\documentclass[draftcls,onecolumn]{IEEEtran}

\usepackage{amsmath,amssymb}
\usepackage[dvips]{graphicx}
\usepackage{cite}
\usepackage[dvipdfm]{hyperref}
\usepackage{subfigure}

\newtheorem{thm}{Theorem}
\newtheorem{de}{Definition}
\newtheorem{claim}{Claim}
\newtheorem{rem}{Remark}

\usepackage{color}

\newcommand{\DEF}{\stackrel{\mbox{\rm \scriptsize def}}{=}}
\newcommand{\pr}{{\rm Pr}}
\renewcommand{\QED}{\hfill $\Box$}

\renewcommand{\t}[1]{\tilde{#1}}

\renewcommand{\c}[1]{{{\mathcal #1}}}

\renewcommand{\o}[1]{{{\overline #1}}}
\newcommand{\sssi}{$(2,2)$--threshold scheme with detectability of impersonation attacks}
\newcommand{\sssc}{threshold scheme with detectability of substitution attacks}

\makeatletter
\def\eqnarray{%
        \stepcounter{equation}%
        \let\@currentlabel=\theequation
        \global\@eqnswtrue\global\@eqcnt\z@
        \tabskip\@centering
        \let\\=\@eqncr
        $$\halign to \displaywidth\bgroup\@eqnsel\hskip\@centering
        $\displaystyle\tabskip\z@{##}$&\global\@eqcnt\@ne
        \hfil$\displaystyle{{}##{}}$\hfil
        &\global\@eqcnt\tw@$\displaystyle\tabskip\z@{##}$\hfil
        \tabskip\@centering&\llap{##}\tabskip\z@\cr}
\makeatother

\begin{document}

\title{Coding Theorems for a $(2,2)$--Threshold Scheme with Detectability of \\Impersonation Attacks\thanks{This paper is presented in part at IEEE International Symposium on Information Theory 2009, and IEEE Information Theory Workshop 2009.}}


\author{Mitsugu~Iwamoto,~\IEEEmembership{Member,~IEEE},
Hiroki~Koga,~\IEEEmembership{Member,~IEEE},\\
        and~Hirosuke~Yamamoto,~\IEEEmembership{Fellow,~IEEE}
\thanks{M.~Iwamoto is with the Center of Frontier Science and Engineering, 
        University of Electro-Communications, 
 1--5--1 Chofugaoka, Chofu-shi, Tokyo 182--8585,
Japan (e-mail: mitsugu@inf.uec.ac.jp).
}
\thanks{H.~Koga is with 
 Tsukuba University, 1--1--1, Tennoudai, Tsukuba-shi, 305--8577 Japan (e-mail: koga@iit.tsukuba.jp). 
}
\thanks{H.~Yamamoto is with 
Graduate School of Frontier Sciences, University of Tokyo, 
 5--1--5 Kashiwanoha, Kashiwa-shi, Chiba 277--8561, Japan 
(e-mail: Hirosuke@ieee.org)}
}\maketitle

\begin{abstract}
In this paper, we discuss coding theorems on a $(2, 2)$--threshold scheme 
in the presence of an opponent who impersonates one of 
the two participants in an asymptotic setup. 
We consider a situation where $n$ secrets $S^n$ from a memoryless source 
is blockwisely encoded to two shares and the two shares are decoded to 
$S^n$ with permitting negligible decoding error. 
We introduce correlation level of the two shares and characterize 
the minimum attainable rates of the shares and a uniform random number 
for realizing a $(2, 2)$--threshold scheme that is secure against 
the impersonation attack by an opponent. It is shown that, 
if the correlation level between the two shares equals to an 
$\ell \ge 0$, the minimum attainable rates coincide with $H(S)+\ell$, 
where $H(S)$ denotes the entropy of the source, and 
the maximum attainable exponent of the success probability 
of the impersonation attack equals to $\ell$. We also give 
a simple construction of an encoder and a decoder using 
an ordinary $(2,2)$--threshold scheme where the two shares are 
correlated and attains all the bounds.  
\end{abstract}


\section{Introduction}

\subsection{Background and Motivations}
\begin{figure*}[tb]
\begin{center}
\subfigure[A $(2,2)$--threshold scheme with an opponent who impersonates a participant who has a share $X$ (impersonation attack)]
{\includegraphics[width=.72\textwidth]{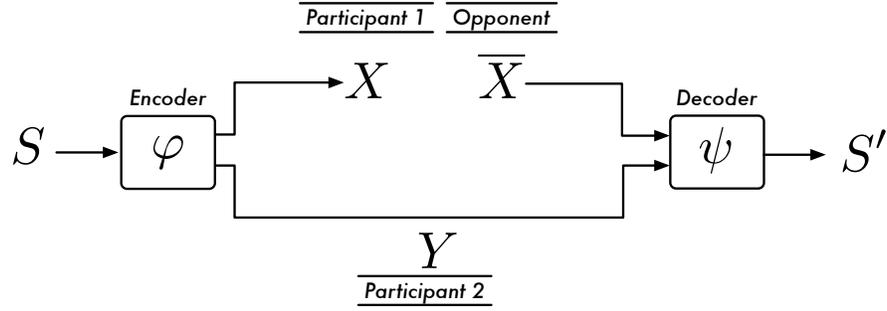}
\label{fig:impersonation}}\\
\subfigure[A $(2,2)$--threshold scheme with an opponent who substitutes a share $\overline{X}$ for $X$ (substitution attack)]
{\includegraphics[width=.72\textwidth]{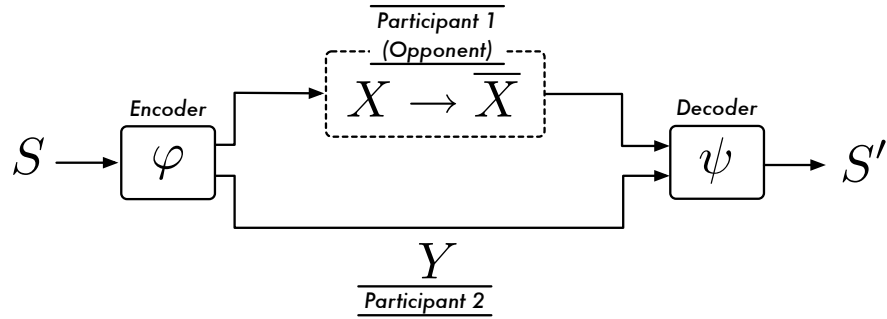}
\label{fig:substitution}}
\end{center}
\caption{Two $(2,2)$--threshold schemes with an opponent}
\label{fig:opponents}
\end{figure*}

\begin{figure*}[tb]
\begin{center}
\includegraphics[width=.88\textwidth]{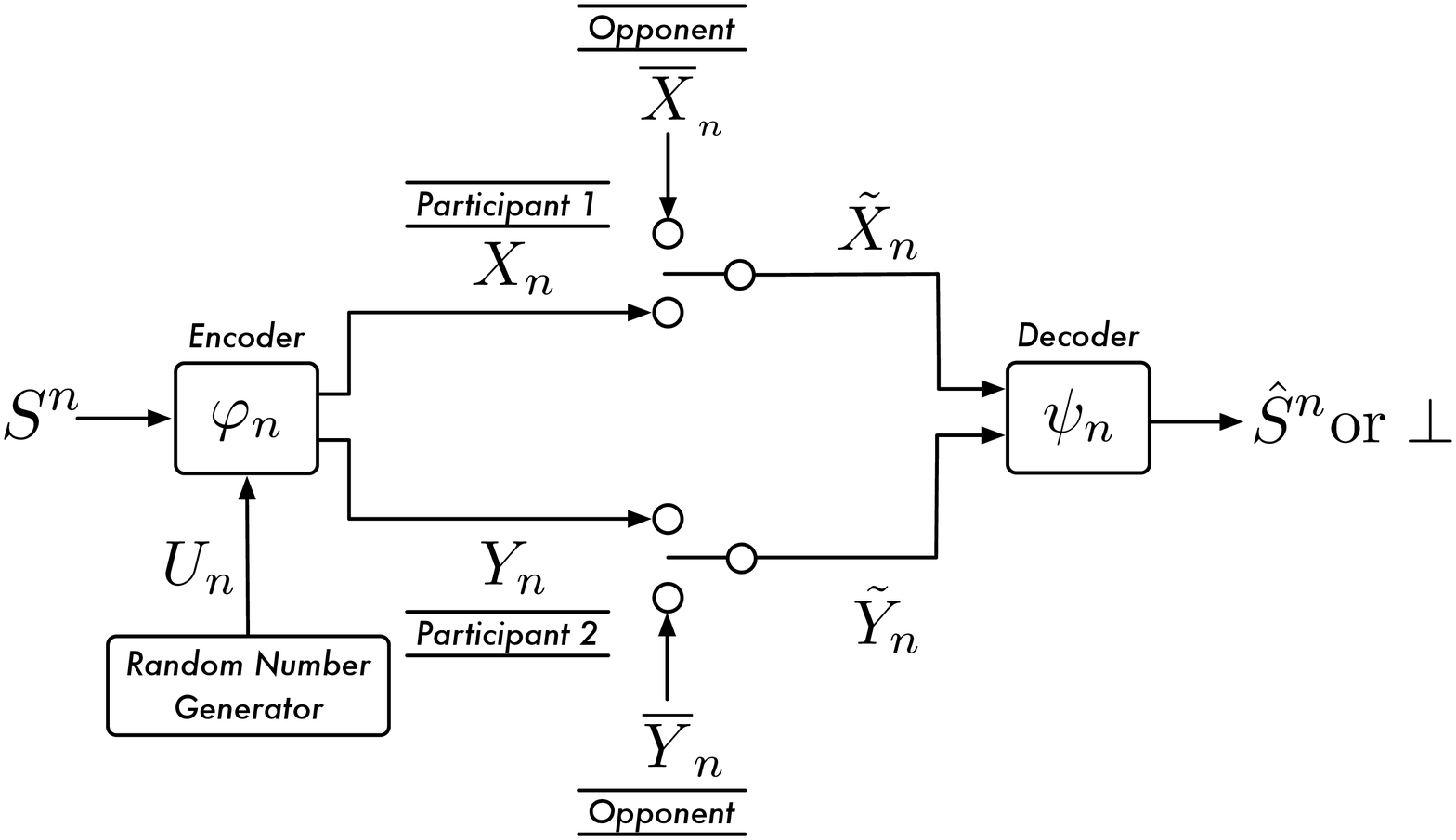}
\end{center}
\caption{A system model of \sssi}
\label{fig:system}
\end{figure*}

A secret sharing scheme \cite{S-cacm79,B-afips79} is a well-known 
cryptographic technique that enables us to share a secret data 
among users. In $(t,m)$--threshold schemes, for example, 
a secret $S$ is encoded to $m$ {\em shares}, and the $m$ shares are 
distributed to respective participants. 
Any $t$ out of $m$ participants can recover $S$, while $t-1$ or fewer participants cannot obtain any information on $S$ in the sense of {\em unconditional security}.

In this paper, we focus on the secret sharing scheme in the presence of opponents. 
The objective of the opponents is cheating honest participants. That is, the opponents forge
their shares and try to cheat the honest participants by injecting the forged shares 
in the recovery phase of $S$. This problem was firstly discussed 
by McEliece-Sarwate \cite{MS-cacm81} and Karnin-Greene-Hellman 
\cite{KGH-it83} from the viewpoint of error-correcting codes. 
In particular, Karnin-Greene-Hellman  \cite{KGH-it83} and Tompa-Woll \cite{TW-jc88} clarified that it is impossible 
to detect cheating in Shamir's secret sharing scheme \cite{S-cacm79}.  
In addition, a construction of a cheating-detectable 
secret sharing scheme is proposed in \cite{TW-jc88} as an extension of Shamir's 
secret sharing scheme although it is much inefficient. 
So far several schemes have been proposed to overcome such 
disadvantages \cite{CSV-ecrypt93,KOO-crypto95,OKS-siam06,OA-acrypt06}. 
In particular, Ogata-Kurosawa-Stinson \cite{OKS-siam06} derived 
a lower bound on sizes of shares under a given maximum success 
probability $\varepsilon$ of cheating and the lower bound is attained 
if and only if a difference set exists.

In cheating-detectable threshold schemes, the shares must satisfy
unforgeablity as well as the ordinary requirements as
a threshold scheme. We can actually consider two types of attacks,
{\em impersonation attacks} and {\em substitution attacks}, similarly to
the attacks against secret-key authentication systems \cite{S-crypto84}.
In the impersonation attack, opponents intend to impersonate
 participants by injecting forged shares without using the 
legitimate shares. The impersonation attack is regarded as
successful if the forged shares are accepted in a recovery phase
of a secret. On the other hand, in the substitution attack,
some of the participants are malicious and forge their shares
by using their shares. The objective of the malicious participants is cheating honest participants who want to recover $S$ from their shares. 

For instance, 
Figure \ref{fig:opponents} shows the two types of attacks against
a $(2,2)$--threshold scheme with two shares $X$ and $Y$.
We assume that the ordinary requirements as a $(2,2)$--threshold schemes,
$H(S|X) = H(S|Y)=H(S)$ and $H(S|XY)=0$, are satisfied.
In Fig.\ \ref{fig:impersonation} an opponent generates a forged share
$\overline{X}$ without using $X$ and $Y$ and
tries to impersonate participant 1 who have a share $X$. In 
Fig.\ \ref{fig:substitution} a participant with $X$ forges $\overline{X}$
by using $X$, but not using $Y$. We assume that in both cases $\overline{X}$ is generated probabilistically.
Then, it is important to notice that,
$\overline{X}$ is independent of $(X,Y)$ in Fig.\ \ref{fig:impersonation},
while $Y$, $X$ and $\overline{X}$ form a Markov chain
in this order in Fig.\ \ref{fig:substitution}. 
Thus, considering the two types of attacks against threshold schemes
corresponds to  giving two kinds of  probabilistic structures
for all the shares including the forged share.

Cheating-detectable secret sharing schemes are usually designed to detect
substitution attacks \cite{TW-jc88,CSV-ecrypt93,KOO-crypto95,OKS-siam06,OA-acrypt06} in a non-asymptotic setup,
i.e., the decoding error is not allowed and the block coding is not
considered. These studies treat the case where a coalition of more than
one malicious participants generates forged shares. However,
there exist the following drawbacks in cheating-detectable
secret sharing schemes:

\if
In the same way as ordinary secret sharing schemes, the above schemes are discussed in a non-asymptotic setup, i.e., the decoding error is not allowed and the block coding is not considered. As a result, there exist the following drawbacks in cheating-detectable secret sharing schemes: 
\fi

\begin{itemize}
\item According to \cite{OKS-siam06}, it is easy to derive  the lower bounds of share rates, i.e., information bits per secret needed to describe shares, under a given success probability of cheating. Unfortunately, however, this result implies that the optimal share rates increase in order at least $1/\varepsilon$ as $\varepsilon \rightarrow 0$, and hence, an arbitrarily small success probability of cheating cannot be realized with fixed finite share rates. 
\item An extension of Shamir's $(t,m)$--threshold scheme in \cite{TW-jc88} can detect both substitution and impersonation attacks. This scheme is simple but inefficient from the viewpoint of share sizes. In addition,  the optimal construction \cite{OKS-siam06} is based on a combinatoric structure called a {\em difference set}, where the difference set exists only in limited cases and therefore restricts sizes of a secret and shares.
 Hence, even in a $(2,2)$--threshold scheme, we cannot apply the optimal scheme to a secret $S$ of  arbitrarily given size.  
\item Almost all constructions include the assumption that a secret is generated subject to a uniform probability distribution. This means that developing a near-optimum cheating-detectable secret sharing scheme for a secret subject to a non-uniform becomes another problem \cite{OA-acrypt06}.  
\end{itemize}

In this paper, we focus on the impersonation attack against a $(2,2)$--threshold scheme. 
Since the impersonation attack is weaker than the substitution attack, 
the impersonation attack is rarely discussed especially in the framework 
of secret sharing schemes. However, if we discuss the threshold scheme 
secure against impersonation attack in a certain asymptotic setup,  
we can unveil another information-theoretic aspect. In fact, we can find 
connections to hypothesis testing, authentication codes and Shannon's cipher system. 
In a practical point of view, we can consider a situation where
impersonation attack seems to be valid. Suppose that in a $(2,2)$--threshold
scheme one of the shares, say $X$, is a uniform random number
that is independent of a secret $S$. In this case, the participants
having $X$ may generate $\overline{X}$ subject to a distribution
close to the uniform distribution because analysis of $X$ gives
almost no information to the participant.

\subsection{Contribution of This Study}
In this paper, we formulate the problem of a threshold scheme secure against impersonation attacks 
in Shannon-theoretic asymptotic setup \cite{Y-it86,K-it06}, and unveil new features included in the problem. 
We consider a situation 
where $n$ secrets that are generated from a discrete memoryless source 
are blockwisely encoded to two shares and the two shares are decoded 
to $n$ secrets with permitting negligible decoding error.
While we consider impersonation attacks, the asymptotic $(2,2)$--threshold scheme treated in this paper has the following features which resolve the three drawbacks  pointed above in cheating-detectable secret sharing schemes:
\begin{itemize}
\item An exponentially small success probability of impersonation attack is realized under finite share rates if the blocklength is sufficiently large. 
\item The scheme uses no combinatoric structure and is applicable to arbitrary size of a secret. 
\item The probability distribution of a secret is arbitrary. In addition, the scheme can be applied to a more general class of sources. 
\end{itemize}

Specifically, we give coding theorems on the $(2,2)$--threshold scheme for two cases of 
blockwise encoding and symbolwise encoding. In both cases we are interested in 
the minimum attainable rates for not only the two shares but also the uniform 
random number needed to a dealer for realizing a cheating-detectable 
$(2,2)$--threshold scheme in an asymptotic sense. We also evaluate the maximum attainable 
exponent of the success probability of the impersonation attack. 
It turns out that, if the two shares are correlated, 
we can easily realize the $(2,2)$--threshold scheme 
in an asymptotic sense that is secure against the impersonation 
attack. This fact motivates us to define a notion of 
{\em correlation level} of the two shares as the limit of 
the normalized mutual information between the two shares. 
In a non-asymptotic setup, we note that correlated shares are firstly discussed in \cite{SV-SIAMdm88} based on a combinatorial argument.

In the case of blockwise encoding, we consider an encoder 
that encodes $n$ secrets $S^n=S_1S_2 \cdots S_n$ blockwisely to 
two shares $X_n$ and $Y_n$ by using a uniform random number $U_n$, 
where throughout the paper the superscript $n$ denotes the length and 
the subscripts $n$ indicate dependency of $n$. The two shares 
$X_n$ and $Y_n$ are decoded to $S^n$ with decoding error probability 
$P_n^e$ that satisfies $P_n^e \to 0$  as $n \to \infty$. 
The two shares are required to satisfy the security criteria 
$I(S^n;X_n) /n \to 0$ and $I(S^n;Y_n)/n \to 0$ 
as $n \to \infty$, where $I(\,\cdot \,;\,\cdot\,)$ denotes the 
mutual information. We can prove that, if the correlation level 
of the shares is equal to $\ell$, none of the rates of 
$X_n,Y_n$ and $U_n$ cannot be less than $H(S)+\ell$, 
where $H(S)$ denotes the entropy of the source, 
and the exponent of the success probability of impersonation attack 
cannot be greater than $\ell$ ({\em converse part}). 
Furthermore, we can prove the existence of a symbolwise of pairs of 
an encoder and a decoder that attains all the bounds shown in the converse part ({\em direct part}). Both the claims of the direct and the converse parts 
are easily extend to the case where $S^n$ is generated from 
a stationary ergodic source. 

In the case of symbolwise encoding, we consider an encoder 
that encodes $n$ secrets $S^n$ 
to two shares $X^n=X_1 X_2\cdots X_n$ and $Y^n=Y_1 Y_2\cdots Y_n$ 
of length $n$ by using $n$ uniform random numbers 
$U^n=U_1 U_2 \cdots U_n$. In fact, $X^n$ and $Y^n$ are generated by 
$(X_i,Y_i) = f(S_i,U_i)$ for $i=1,2,\ldots,n$, 
where $f$ is an arbitrary deterministic encoder of 
an ordinary $(2,2)$--threshold scheme satisfying 
$H(S_i|X_i)=H(S_i|Y_i)=H(S_i)$ and $H(S_i|X_iY_i) = 0$. 
Denote by $g$ a deterministic map satisfying $S_i = g(X_i,Y_i)$. 
We choose an appropriate $f$ so that 
$(X_i,Y_i),\,i=1,2,\ldots,n$, can be regarded as i.i.d. correlated 
random variables. It is shown that we can realize 
a $(2,2)$--threshold scheme in an asymptotic sense 
in which $P_n^e$ vanishes as $n \to \infty$,  
$X^n$ and $Y^n$ satisfy a stronger requirement on the secrecy 
$I(S^n;X^n)=I(S^n;Y^n)=0$ and the exponent of the success probability 
of the impersonation attack is optimal. In the proof we construct 
a decoder of $X^n$ and $Y^n$ by using $g$ and a one-sided test 
for verifying the joint typicality of $X^n$ and $Y^n$. 
This kind of symbolwise setup is first discussed 
in \cite{KY-ieice00} for authentication code. 

\subsection{Related Works, Organization}

The $(2,2)$--threshold scheme secure against the impersonation attack
is motivated from the Shannon-theoretic authentication codes
\cite{S-crypto84,KY-ieice00,K-ieice00,M-it00}. In particular, in \cite{KY-ieice00} 
the authors discuss
the maximum attainable error exponent on the success probability
of the impersonation attack subject to the vanishing
decoding error probability. However, the results given in this paper
is more involved. In fact, in the framework of $(2,2)$--threshold schemes
we need to guarantee secrecy of a secret given one of the two shares.
In addition, in this paper we succeeded in obtaining not only
such a maximum exponent but also the minimum attainable sizes of the
shares and the uniform random number.

The {\sssi} 
with the blockwise encoder can be viewed as one version of Shannon's 
cipher system (\cite{S-milestone,Hel-it77,Yam-it94,Yam-ieice91,Mau-it93} etc.) 
when one of the shares is an  output from a random number 
generator. In a simple asymptotic setup of 
Shannon's cipher system \cite{Yam-ieice91}, $n$ plaintexts $S^n$ 
generated from a memoryless source are encrypted 
to a cryptogram $W_n$ under a key $U_n$ and 
$W_n$ is decrypted to $S^n$ under the same key $U_n$ 
with permitting decoding error probability $P_n^e$.  
The encoder and the decoder are required to satisfy 
$P_n^e \to 0$ and $I(S^n;W_n)/n \to 0$ as $n \to \infty$. 
In this setup, the minimum attainable rates of the cryptogram and 
the key coincide with the entropy $H(S)$ of the plaintext. 
The coding theorems given in this paper imply the same result 
under an additional requirement such that the correlation level 
of $W_n$ and $U_n$ is equal to zero, i.e., $\ell=0$. 

The {\sssi} with 
the symbolwise encoder is related to the problem of secret key agreement 
\cite{Mau-it93,AC-it93}. 
In the secret key agreement problem of the source type model 
\cite{AC-it93}, two users have $n$ outputs 
$X^n=X_1 X_2 \cdots X_n \in \c{X}^n$ and 
$Y^n=Y_1 Y_2 \cdots Y_n \in \c{Y}^n$ 
from two correlated memoryless source, respectively, 
where $(X_i,Y_i),\,i=1,2,\ldots,n$ are i.i.d. copies of 
$(X,Y) \in \c{X}\times \c{Y}$ subject to a joint probability 
distribution $P_{XY}$. 
The two user try to share a nearly uniform random number 
with the maximum rate $I(X;Y)$ by public communications. 
On the contrary, the symbolwise encoder in the {\sssi} 
can be interpreted as a generator 
of correlated random variables 
$X^n = X_1 X_2 \cdots X_n \in \c{X}^n$ and 
$Y^n = Y_1 Y_2 \cdots Y_n \in \c{Y}^n$ 
given independent random variables $S^n$ and $U^n$, where 
$(X_i,Y_i),\,i=1,2,\ldots,n$, are regarded as $n$ i.i.d. copies of 
$(X,Y) \sim P_{XY}$. 
Since the correlation level $X^n$ and $Y^n$ coincides with $I(X;Y)$, 
the minimum attainable rate of $U^n$ turns out to be 
$H(S)+I(X;Y)$. That is, we need an extra cost of $I(X;Y)$ 
in order to generate correlated two shares. 

The rest of this paper is organized as follows: 
In Section \ref{sec:preliminary}, a 
{\sssi} with correlation level $\ell$ 
in an asymptotic setup is formulated. The coding theorems 
for the blockwise encoder are given in Section \ref{sec:main}. 
Section \ref{sec:proofs} is devoted to the proofs of the 
coding theorems. A construction of encoders and decoders based on 
a non-asymptotic $(2,2)$--threshold scheme 
and its optimality are discussed in Section \ref{sec:simpler}.

\section{Problem Setting}\label{sect:PS}
\label{sec:preliminary}
We consider a $(2,2)$--threshold scheme depicted in Fig.\ \ref{fig:system}. Assume for an integer $n \ge 1$ that a source generates an $n$--tuple of secrets $S^n=S_1S_2\cdots S_n$ independently subject to a probability distribution $P_S$ on a finite set $\c{S}$. Denote by $P_{S^n}$ the probability distribution of $S^n$ induced by $P_S$, and let $P_{S^n}(s^n)$ be the probability that $S^n = s^n$ for an $s^n \in \c{S}^n$. Since the source is memoryless, it holds that $P_{S^n}(s^n)=\prod_{i=1}^nP_{S}(s_i)$ for all $n \ge 1$ where $s^n = s_1s_2\cdots s_n$. 

In Fig.\ \ref{fig:system}, let $U_n$ be the random variable subject to the uniform distribution on a finite set $\c{U}_n$. Assume that $U_n$ is independent of $S^n$. In this paper, we use the subscript $n$ to indicate dependency of $n$, while the superscript $n$ implies the length. We denote by $P_{U_n}$ a probability distribution of $U_n$, i.e., it holds that $P_{U_n}(u_n) = 1/|\c{U}_n|$ for all $u_n  \in \c{U}_n$ where $|\cdot|$ denotes the cardinality.

An encoder is defined as a deterministic map $\varphi_n: \c{S}^n \times \c{U}_n \rightarrow \c{X}_n \times \c{Y}_n$, where $\c{X}_n$ and $\c{Y}_n$ are finite sets in which shares $X_n$ and $Y_n$ take values, respectively.  Hence, we can write
\begin{eqnarray}\label{eq:enc_phi}
(X_n,Y_n)=\varphi_n(S^n,U_n)
\end{eqnarray}
from which we can see that $X_n$ and $Y_n$ are also random variables. The joint probability distribution $P_{X_nY_n}$ of $X_n$ and $Y_n$ is induced from (\ref{eq:enc_phi}). The shares $X_n$ and $Y_n$ are distributed securely to participants 1 and 2, respectively.

Next, consider a situation where an opponent may impersonate one of the two participants. When the opponent impersonates participant 1, the opponent behaves as if he/she were a participant 1 by injecting a forged share $\overline{X}_n \in \c{X}_n$  instead of $X_n$. This attack is regarded as successful if a decoder fails to detect impersonation attacks and outputs an element of $\c{S}^n$ from $\overline{X}_n$ and $Y_n$. Here, we assume that the opponent generates $\overline{X}_n$ {\em without using} $X_n$. According to \cite{S-crypto84,KY-ieice00,K-ieice00,M-it00}, such attack is called {\em impersonation attack} as opposed to substitution attack. Similarly, in the case of deceiving participant 1, the opponent forges a share $\overline{Y}_n$ without using $Y_n$, and tries to impersonate participant 2. In this case, the attack succeeds when the decoder outputs an element of $\c{S}^n$ from $X_n$ and $\overline{Y}_n$. Summarizing, letting $\t{X}_n$ and $\t{Y}_n$ be the inputs to a decoder, the following three cases must be considered: 
\begin{description}
\item[\sf (a0)] $(\t{X}_n,\t{Y}_n) = (X_n,Y_n)$
\item[\sf (a1)] $(\t{X}_n,\t{Y}_n) = (\o{X}_n,Y_n)$
\item[\sf (a2)] $(\t{X}_n,\t{Y}_n) = (X_n,\o{Y}_n)$
\end{description}

A decoder is defined as a deterministic map $\psi_n: \c{X}_n \times \c{Y}_n \rightarrow \c{S}^n \cup \{\perp\}$, where $\perp$ is a symbol to declare the detection of an impersonated attack, i.e., {\sf (a1)} or {\sf (a2)}. We note here that the decoder cannot know  in advance which one of {\sf (a0)}--{\sf (a2)} actually occurs. 
On the other hand, we assume that the opponent knows everything about the encoder and the decoder  except for realizations of $S^n, U_n, X_n$ and $Y_n$.

In this situation, we define success probabilities of impersonation attacks. Let $\c{A}_n \subset \c{X}_n \times \c{Y}_n$ be the region that the decoder $\psi_n$ accepts the pair of shares $(\t{X}_n, \t{Y}_n)$ and outputs an element of $\c{S}^n$, i.e., 
\begin{eqnarray}
\c{A}_n = \{
(x_n,y_n) \in \c{X}_n \times \c{Y}_n : \psi_n(x_n, y_n) \in \c{S}^n\}. 
\end{eqnarray}
Now, recall that the impersonation attack succeeds if the decoder outputs an element of $\c{S}^n$ when one of {\sf (a1)} and {\sf (a2)} occurs. In the case of {\sf (a1)}, i.e., the opponent impersonates participant 1, we note that he/she generates a forged share $\o{X}_n$ according to a probability distribution $P_{\o{X}_n}$ independently from $S^n$, $U_n$, $X_n$, and $Y_n$. In addition, the opponent tries to optimize $P_{\o{X}_n}$ so that $(\o{X}_n,Y_n)$ can be accepted by the decoder with the maximum probability. This motivates us to define a success probability to impersonate participant 1 by 
\begin{eqnarray}\label{eq:succ_prob_X}
 P_n^{X} &=& \max_{P_{\o{X}_n}}\pr \{(\o{X}_n,Y_n) \in \c{A}_n\}
\end{eqnarray}
where the maximization of $P_{\o{X}_n}$ is taken over all probability distributions on $\c{X}_n$, and $\pr\{\cdot\}$ means the probability with respect to the (joint) probability distribution of random variable(s) between the parentheses, i.e., $(\o{X_n},Y_n) \sim P_{\o{X_n}Y_n}=P_{\o{X}_n} P_{Y_n}$ in this case. Similarly, the maximum success probability for the impersonation to participant $2$ can be defined as 
\begin{eqnarray}\label{eq:succ_prob_Y}
 P_n^{Y} &=& \max_{P_{\o{Y}_n}}\pr \{(X_n,\o{Y}_n) \in \c{A}_n\}
\end{eqnarray}
where $\pr \{ \cdot \}$ is taken with respect to $(X_n, \o{Y}_n) \sim P_{X_n\o{Y}_n}=P_{X_n}P_{\o{Y}_n}$. 

The decoding error occurs when $S^n$ is not correctly decoded from legitimate shares in the case of {\sf (a0)}. Hence, the decoding error probability can be written as 
\begin{eqnarray}
P_n^e = \pr \{\psi_n(\varphi_n(S^n,U_n)) \neq S^n\}. 
\end{eqnarray}
It is easy to see that if $(x_n,y_n) \not\in \c{A}_n$, then $\psi_n(x_n,y_n) = \bot \not\in \c{S}^n$. Hence, we have
\begin{eqnarray}\label{eq:error}
P_n^e \ge \pr \{(X_n,Y_n) \not\in \c{A}_n\}
\end{eqnarray}
for any pair of an encoder $\varphi_n$ and a decoder $\psi_n$.

Now, we can define a $(2,2)$--threshold scheme in an asymptotic setup as follows:

\bigskip
\begin{de}\label{def:main}
We say that a sequence $\{(\varphi_n,\psi_n)\}_{n=1}^\infty$ of an encoder $\varphi_n$ and a decoder $\psi_n$ asymptotically realizes a $(2,2)$--threshold scheme if it satisfies
\begin{eqnarray}\label{eq:asym_err}
\lim_{n \rightarrow \infty} P_n^e = 0
\end{eqnarray}
and
\begin{eqnarray}\label{eq:security}
\lim_{n \rightarrow \infty}\frac{1}{n} I(S^n;X_n) = \lim_{n \rightarrow \infty}\frac{1}{n}I(S^n;Y_n) = 0
\end{eqnarray}
where $I(\cdot;\cdot)$ denotes the mutual information. 
\end{de}
\bigskip

The condition  (\ref{eq:asym_err}) guarantees that the decoding error probability is negligible if the blocklength $n$ is sufficiently large. Note that Fano's inequality \cite[Theorem 2.10.1]{CT-06} tells us that
\begin{eqnarray}\label{eq:fano}
\frac{1}{n}H(S^n|X_nY_n) \le \frac{1}{n}h(P_n^e) + P_n^e \log |\c{S}| 
\end{eqnarray}
where $\log(\cdot) = \log_2(\cdot)$ throughout the paper, and $H(\cdot|\cdot)$ and $h(\cdot)$ are the conditional and the binary entropies, respectively. Hence, if (\ref{eq:asym_err}) is satisfied, then we have 
\begin{eqnarray}\label{eq:soundness2}
\lim_{n \rightarrow \infty} \frac{1}{n} H(S^n|X_nY_n) =0 
\end{eqnarray}
due to the non-negativity of the conditional entropy. On the other hand, the condition (\ref{eq:security}) ensures that $S^n$ is secure  against the leakage from one of $X_n$ and $Y_n$ if $n$ is sufficiently large. That is, $S^n$ and either one of the shares are almost independent under such a condition. We also note that, since $S^n$ is generated from a memoryless source, (\ref{eq:security})  implies that
\begin{eqnarray}\label{eq:security2}
\lim_{n \rightarrow \infty} \frac{1}{n} H(S^n|X_n) =
\lim_{n \rightarrow \infty} \frac{1}{n} H(S^n|Y_n) = H(S) 
\end{eqnarray}
where $H(\cdot)$ denotes the entropy. 

We conclude this section with introducing a notion of {\em correlation level}. The mutual information of two shares plays a crucial role in detecting impersonation attacks, which will be clarified in the following sections. 
\bigskip
\begin{de} 
Let $\{(X_n,Y_n)\}_{n = 1}^\infty$ be a pair of shares generated by a sequence of encoders $\{\varphi_n\}_{n = 1}^\infty$. Then, a non-negative number $\ell$ is said to be a correlation level of 
$\{(X_n,Y_n)\}_{n = 0}^\infty$ if it holds that 
\begin{eqnarray}\label{eq:corr-level}
\lim_{n \rightarrow \infty} \frac{1}{n} I(X_n;Y_n) = \ell.
\end{eqnarray}
In particular, if a sequence $\{(\varphi_n,\psi_n)\}_{n = 1}^\infty$ of an encoder $\varphi_n$ and a decoder $\psi_n$ satisfies Definition \ref{def:main} and the sequence of shares $\{(X_n,Y_n)\}_{n = 0}^\infty$ generated from $\{\varphi_n\}_{n =1}^\infty$ satisfies (\ref{eq:corr-level}), we say that $\{(\varphi_n,\psi_n)\}_{n = 1}^\infty$ asymptotically realizes a $(2,2)$--threshold scheme with correlation level $\ell$. 
\end{de}
\bigskip
\begin{rem}
Note that the sequence $\{I(X_n;Y_n)/n\}_{n=1}^\infty$  in (\ref{eq:corr-level}) does not have a limit in general if $\{(X_n,Y_n)\}_{n = 1}^\infty$ is generated by an arbitrary sequence of encoders $\{\varphi_n\}_{n=1}^\infty$. Hence, (\ref{eq:corr-level}) actually requires the existence of the limit for the sequence $\{I(X_n;Y_n)/n\}_{n=1}^\infty$, and the limit equals to $\ell$. 
\end{rem}

\section{Coding Theorems for a $(2,2)$--Threshold Scheme with Detectability of Impersonation Attacks}\label{sec:main}

In this section, we give coding theorems for $\{(\varphi_n,\psi_n)\}_{n=1}^\infty$ that asymptotically realizes a $(2,2)$--threshold scheme with correlation level $\ell$. We are interested in not only the rates of $X_n$, $Y_n$ and $U_n$ but also the exponents of $P_n^X$ and $P_n^Y$ of the sequence $\{(\varphi_n,\psi_n)\}_{n=1}^\infty$. The following theorem is the converse part of the coding theorem with respect to such rates and exponents. 

\bigskip
\begin{thm}\label{thm:converse}
For any sequence  $\{(\varphi_n,\psi_n)\}_{n=1}^\infty$ of an encoder $\varphi_n$ and a decoder $\psi_n$ that asymptotically realizes a $(2,2)$--threshold scheme with correlation level $\ell$, it holds that
\begin{eqnarray}
\label{eq:th1-1}
\liminf_{n \rightarrow \infty}\frac{1}{n}\log|\c{X}_n| &\ge& H(S) + \ell \\
\label{eq:th1-2}
\liminf_{n \rightarrow \infty}\frac{1}{n}\log|\c{Y}_n| &\ge& H(S) + \ell \\
\label{eq:th1-3}
\liminf_{n \rightarrow \infty}\frac{1}{n}\log|\c{U}_n| &\ge& H(S) + \ell 
\end{eqnarray} 
and 
\begin{eqnarray}
\label{eq:th1-4}
\limsup_{n \rightarrow \infty}\max \left\{-\frac{1}{n} \log  P_n^X, -\frac{1}{n} \log  P_n^Y \right\} \le \ell.
\end{eqnarray}
\end{thm}

\bigskip
Theorem \ref{thm:converse} is proved in Section \ref{sec:converse}. Theorem \ref{thm:converse} tells us that for an arbitrarily small $\gamma > 0$ the rates of $X_n$, $Y_n$ and $U_n$ cannot be  less than $H(S)+\ell-\gamma$ for all sufficiently large $n$, where $\ell$ is an arbitrarily given correlation level. In fact, by noticing that $H(S) + \ell \ge H(S)$ for any $\ell \ge 0$, the bounds on the right hand sides of (\ref{eq:th1-1})--(\ref{eq:th1-3}) coincide with the bounds in \cite[Theorem 1]{K-it06} for $(2,2)$--threshold schemes when $\ell = 0$. Theorem \ref{thm:converse} also indicates that the correlation level of shares is an upper bound on the exponents of $ P_n^X$ and $ P_n^Y$. 

The direct part of the coding theorem corresponding to Theorem \ref{thm:converse} is as follows:

\bigskip
\begin{thm}\label{thm:direct}
For an arbitrarily given non-negative number $\ell \ge 0$, there exists a sequence $\{(\varphi_n^*,\psi_n^*)\}_{n = 1}^\infty$ of an encoder $\varphi_n^*$ and a decoder $\psi_n^*$ that asymptotically realizes a $(2,2)$--threshold scheme with correlation level $\ell$ satisfying 
\begin{eqnarray}
\label{eq:direct1}
\limsup_{n \rightarrow \infty} \frac{1}{n} \log |\c{X}_n| \le H(S) + \ell \\
\label{eq:direct2}
\limsup_{n \rightarrow \infty} \frac{1}{n} \log |\c{Y}_n| \le H(S) + \ell \\
\label{eq:direct3}
\limsup_{n \rightarrow \infty} \frac{1}{n} \log |\c{U}_n| \le H(S) + \ell
\end{eqnarray}
and
\begin{eqnarray}
\label{eq:direct4}
\liminf_{n \rightarrow \infty} \min\left\{ 
-\frac{1}{n} \log  P^X_n, -\frac{1}{n} \log  P^Y_n
\right\} \ge \ell.
\end{eqnarray}
In particular, the above $\{(\varphi_n^*,\psi_n^*)\}_{n = 1}^\infty$ also satisfies
\begin{eqnarray}
\label{eq:direct5}
I(S^n;X_n) = I(S^n;Y_n) = 0 ~~\mbox{for all}~~n \ge 1
\end{eqnarray}
which is stronger than the condition in (\ref{eq:security}). 
\end{thm}
\bigskip

The proof of Theorem \ref{thm:direct} is given in Section \ref{sec:direct}.
\bigskip
\begin{rem}
Theorem \ref{thm:converse} guarantees that 
$\{(\varphi_n^*,\psi_n^*)\}_{n = 1}^\infty$
in Theorem \ref{thm:direct} attains the minimum rates of $X_n$, $Y_n$, and $U_n$, and the maximum exponents of $ P_n^X$ and $ P_n^Y$.  Furthermore, the limits exist for these rates and exponents, i.e., it holds that 
\begin{eqnarray}
\lim_{n \rightarrow \infty} \frac{1}{n} \log |\c{X}_n| = 
\lim_{n \rightarrow \infty} \frac{1}{n} \log |\c{Y}_n| = 
\lim_{n \rightarrow \infty} \frac{1}{n} \log |\c{U}_n| = H(S) + \ell
\end{eqnarray}
and 
\begin{eqnarray}
\lim_{n \rightarrow \infty} -\frac{1}{n} \log  P_n^X =
\lim_{n \rightarrow \infty} -\frac{1}{n} \log  P_n^Y = \ell.
\end{eqnarray}
\end{rem}
\bigskip

\section{Proofs of Theorems \ref{thm:converse} and \ref{thm:direct}}\label{sec:proofs}
This section is devoted to the proofs of Theorems  \ref{thm:converse} and \ref{thm:direct}. In the proof of Theorem \ref{thm:converse}, we use a relationship between hypothesis testing and the {\sssi}
 with correlation level $\ell$, which originates from \cite{M-it00} and developed by \cite{KY-ieice00,K-ieice00}. 
\subsection{Proof of Theorem \ref{thm:converse}}\label{sec:converse}
Fix $\ell \ge 0$ arbitrarily. We first prove (\ref{eq:th1-1}). From the basic properties of the entropy and the mutual information, it holds that
\begin{eqnarray}
\nonumber
H(X_n) &=& I(X_n;Y_n) + H(X_n|Y_n) \\
\nonumber
&\ge& 
 I(X_n;Y_n) + H(X_n|Y_n) - H(X_n|Y_nS^n)\\
\nonumber
 &=& I(X_n;Y_n) + I(X_n;S^n|Y_n)\\
 &=& 
I(X_n;Y_n) + H(S^n|Y_n) - H(S^n|X_nY_n).
\label{eq:inequality}
\end{eqnarray} 
Hence, (\ref{eq:th1-1}) is established because 
\begin{eqnarray}
\nonumber
\liminf_{n \rightarrow \infty} \frac{1}{n} \log|\c{X}_n| &\ge &
\liminf_{n \rightarrow \infty} \frac{1}{n} H(X_n) \\
\nonumber
&\ge &
\liminf_{n \rightarrow \infty} \frac{1}{n} I(X_n;Y_n) + \liminf_{n \rightarrow \infty} \frac{1}{n}H(S^n|Y_n) - \limsup_{n \rightarrow \infty} \frac{1}{n}H(S^n|X_nY_n)\\
\label{eq:eval_X}
&=& \ell  + H(S) 
\end{eqnarray}
where the last inequality and the equality are due to (\ref{eq:inequality}) and (\ref{eq:soundness2})--(\ref{eq:corr-level}), respectively. We can establish (\ref{eq:th1-2}) in essentially the same way.

Next, we prove (\ref{eq:th1-3}). Since the encoder $\varphi_n$ is deterministic for each $n \ge 1$, we have
\begin{eqnarray}
\nonumber
H(X_nY_n) &\le& H(S^nU_n)\\
\label{eq:30}
&=& nH(S) + H(U_n)
\end{eqnarray}
for all $n \ge 1$, where the equality follows because $S^n$ is independent of $U_n$ and is generated from a memoryless source. On the other hand, recalling that 
\begin{eqnarray}\label{eq:31}
H(X_nY_n) = H(X_n) + H(Y_n) - I(X_n;Y_n)
\end{eqnarray}
it follows from (\ref{eq:30}) and (\ref{eq:31}) that 
\begin{eqnarray}
\nonumber
\frac{1}{n} \log|\c{U}_n| &=& \frac{1}{n} H(U_n) \\
\nonumber
&\ge& \frac{1}{n} H(X_nY_n) - H(S)\\
&\ge& \frac{1}{n} \{ H(X_n) + H(Y_n) - I(X_n;Y_n) \} - H(S) 
\end{eqnarray}
for all $n \ge 1$, where the first equality follows from the uniformity of $U_n \in \c{U}_n$. Therefore, we have 
\begin{eqnarray}
\nonumber
\liminf_{n \rightarrow \infty} \frac{1}{n} \log |\c{U}_n| 
& \ge & 
\liminf_{n \rightarrow \infty}  \frac{1}{n} H(X_n) + 
\liminf_{n \rightarrow \infty}  \frac{1}{n} H(Y_n) - 
\limsup_{n \rightarrow \infty}  \frac{1}{n} I(X_n;Y_n) - H(S)\\
\label{eq:eval_S}
&\ge&  H(S) + \ell 
\end{eqnarray}
where the last inequality follows from (\ref{eq:corr-level}) and (\ref{eq:eval_X}). Note that we have $\liminf_{n \rightarrow \infty} H(Y_n) \ge H(S) + \ell$ in the same way as (\ref{eq:eval_X}). 


To prove  (\ref{eq:th1-4}), we use the fact that the decoding error probability and the success probabilities of impersonation attack in a $(2,2)$--threshold scheme with correlation level $\ell$ are closely related to the error probabilities of the first kind and the second kind in hypothesis testing, respectively, which is pointed out in \cite{KY-ieice00,K-ieice00,M-it00}. Let us consider a simple hypothesis test with the following two hypotheses:
\begin{eqnarray}
&& H_0 : (\t{X}_n,\t{Y}_n) \sim P_{X_nY_n}\\
&& H_1 : (\t{X}_n,\t{Y}_n) \sim P_{X_n}P_{Y_n}. 
\end{eqnarray}
Let $\c{A}_n\subset \c{X}_n \times \c{Y}_n$ denote an acceptance region for the null hypothesis $H_0$. Then, the error probability of the first kind and the error probability of the second kind of the above hypothesis testing are given by 
\begin{eqnarray}
\alpha_n &\DEF & \sum_{({x}_n,{y}_n) \in \c{A}_n^c}P_{X_nY_n}({x}_n,{y}_n)
=\pr \left\{(X_n,Y_n) \not\in \c{A}_n
\right\} \\
\beta_n  &\DEF & \sum_{({x}_n,{y}_n) \in \c{A}_n    }P_{X_n}({x}_n)P_{Y_n}({y}_n)
\end{eqnarray}
where $\c{A}_n^c$ denotes the complement set of $\c{A}_n$. It is easy to see from (\ref{eq:error}) that $P_n^e \ge \alpha_n$ holds for any $n \ge 1$. Hence,  in view of (\ref{eq:asym_err}), we have   
\begin{eqnarray}
\label{eq:error-1st-kind}
\lim_{n \rightarrow \infty} \alpha_n = 0.
\end{eqnarray}
Furthermore, it follows from (\ref{eq:succ_prob_X}) that 
\begin{eqnarray}
\nonumber
P_n^{X} &=& \max_{P_{\o{X}_n}}\pr \{(\o{X}_n,Y_n) \in \c{A}_n\}\\
\nonumber
&=&\max_{P_{\o{X}_n}}\sum_{({x}_n,{y}_n) \in \c{A}_n} P_{\o{X}_n}({x}_n)P_{Y_n}({y}_n)\\
&\ge& \beta_n.
\end{eqnarray}
Similarly, we also have $ P_n^{Y} \ge \beta_n$. Therefore, it holds that 
\begin{eqnarray}
\label{eq:error-2nd-kind}
-\frac{1}{n} \log\beta_n \ge
\max\left\{-\frac{1}{n} \log  P_n^X,-\frac{1}{n} \log P_n^Y\right\}~~\mbox{for all}~~n\ge1.
\end{eqnarray}
According to \cite[Theorem 4.4.1]{blahut} and \cite[Theorem 2]{KY-ieice00}, we have 
\begin{eqnarray}
\nonumber
&&I(X_n;Y_n) = \sum_{(x_n,y_n) \atop \in \c{X}_n \times \c{Y}_n} P_{X_nY_n}({x}_n,{y}_n) \log \frac{P_{X_nY_n}({x}_n,{y}_n)}{P_{X_n}({y}_n)P_{Y_n}({y}_n)}\\
\nonumber
&&= \sum_{(x_n,y_n)  \in \c{A}_n} P_{X_nY_n}({x}_n,{y}_n) \log \frac{P_{X_nY_n}({x}_n,{y}_n)}{P_{X_n}({x}_n)P_{Y_n}({y}_n)}+\sum_{(x_n,y_n)  \in \c{A}_n^c} P_{X_nY_n}({x}_n,{y}_n) \log \frac{P_{X_nY_n}({x}_n,{y}_n)}{P_{X_n}({x}_n)P_{Y_n}({y}_n)}\\
\nonumber
&&\ge\left(\sum_{\c{A}_n}P_{X_nY_n}({x}_n,{y}_n) \right)\log 
\frac{\sum_{\c{A}_n}P_{X_nY_n}({x}_n,{y}_n) }{\sum_{\c{A}_n} P_{X_n}({x}_n)P_{Y_n}({y}_n)}+\left(\sum_{\c{A}^c_n}P_{X_nY_n}({x}_n,{y}_n)  \right)\log \frac{\sum_{\c{A}^c_n} P_{X_nY_n}({x}_n,{y}_n) }{\sum_{\c{A}_n^c} P_{X_n}({x}_n)P_{Y_n}({y}_n)}\\
\nonumber
&&= (1-\alpha_n) \log \frac{1-\alpha_n}{\beta_n} + \alpha_n \log \frac{\alpha_n}{1-\beta_n}\\
\nonumber
&&= -h(\alpha_n) - (1 - \alpha_n) \log \beta_n - \alpha_n\log (1 - \beta_n)\\
&&\ge -h(\alpha_n) - (1 - \alpha_n) \log \beta_n
\label{eq:eval1}
\end{eqnarray}
where the first inequality follows from the log sum inequality, and the second inequality holds because $-\alpha_n \log(1-\beta_n) \ge 0$. Hence, it follows from  (\ref{eq:error-2nd-kind}) and (\ref{eq:eval1})  that 
\begin{eqnarray}\label{eq:eval2}
\frac{1}{n}I(X_n;Y_n) \ge -\frac{h(\alpha_n)}{n} + (1 - \alpha_n)
\max \left\{
-\frac{1}{n} \log  P_n^X,- \frac{1}{n} \log P_n^Y\right\}~~\mbox{for all}~~n \ge 1. 
\end{eqnarray}
Therefore, we have (\ref{eq:th1-4}) by taking the limit superior of both sides of (\ref{eq:eval2}) and noticing (\ref{eq:error-1st-kind}). \QED
\bigskip
\begin{rem}
The claim of Theorem \ref{thm:converse} can be easily extended to the case 
where $S^n$ is generated from a stationary source. 
For the case of the stationary source, 
the entropy $H(S)$ in the statement of Theorem~\ref{thm:converse} is replaced with 
the entropy rate $H \DEF \lim_{n \to \infty} H(S^n)/n$. 
By recalling the existence of the limit of 
$\{H(S^n)/n\}_{n=1}^\infty$ \cite[Theorem 4.2.1]{CT-06}, we can easily check 
that both the left hand sides of (\ref{eq:eval_X}) and (\ref{eq:eval_S}) are bounded 
by $H + \ell$. 
\end{rem}

\subsection{Proof of Theorem \ref{thm:direct}} \label{sec:direct}
We choose arbitrarily a sequence $\{\gamma_n\}_{n=1}^\infty$ of positive numbers that satisfies $\lim_{n\rightarrow\infty}\gamma_n=0$ and $\lim_{n \rightarrow \infty} \sqrt{n}\gamma_n = \infty$. Let $\c{T}_{\gamma_n}$ be the typical set defined by 
\begin{eqnarray}
\label{eq:typical}
\c{T}_{\gamma_n}=
\left\{
s^n \in \c{S}^n: \left| \frac{1}{n} \log \frac{1}{P_{S^n}(s^n)} -H(S)\right| \le \gamma_n
\right\}. 
\end{eqnarray}
Then, it is well-known that (e.g., see \cite[Theorem 3.1.2]{CT-06}) ${\cal T}_{\gamma_n}$ satisfies the following properties: 
\begin{eqnarray}
\label{eq:lln}
&&\lim_{n \rightarrow \infty} \pr\{S^n \in \c{T}_{\gamma_n}\} = 1\\
\label{eq:type}
&&|\c{T}_{\gamma_n}|  \le  2^{n \{H(S)+ \gamma_n\}}~~\mbox{for all}~~n \ge 1.
\end{eqnarray} 
For an arbitrary $\ell \ge 0$, let $\c{L}_n \DEF \{0,1,\ldots,L_n-1\}$ and $\c{M}_n \DEF \{0,1,\ldots, M_n-1\}$ be sets of integers where $L_n \DEF \lfloor 2^{n\ell} \rfloor$ and $M_n \DEF |\c{T}_{\gamma_n}|$.

In the following, we construct a sequence $\{(\varphi^*_n,\psi_n^*)\}_{n=1}^\infty$ of an encoder $\varphi_n^*$ and a decoder $\psi_n^*$ that asymptotically realizes a $(2,2)$--threshold scheme with correlation level $\ell$ satisfying $|\c{X}_n| = |\c{Y}_n| = |\c{U}_n| = L_n(M_n+1)$. 

The encoder $\varphi_n^*$ can be constructed as follows: Since $M_n = |\c{T}_{\gamma_n}|$, there exists a bijection $\xi_n: \c{T}_{\gamma_n} \rightarrow \c{M}_n$. Furthermore, define a map $\xi_n^+: \c{S}^n \rightarrow \c{M}_n^+$ where $\c{M}_n^+\DEF\c{M}_n \cup \{M_n\}$ by 
\begin{eqnarray}
\xi^+_n(s^n) =
\left\{
\begin{array}{cllll}
\xi_n (s^n), & \mbox{if}~s^n \in \c{T}_{\gamma_n}\\
M_n, & \mbox{otherwise}
\end{array}
\right.
\end{eqnarray}
and let $Z_n \DEF \xi_n^+(S^n)$. Denote by $U_n^{\c{L}}$ and $U_n^{\c{M}}$ the random variables subject to the uniform distribution on $\c{L}_n$ and $\c{M}_n^+$, respectively, and define $U_n = (U_n^\c{L},U_n^\c{M})$. In addition, we define two shares by 
\begin{eqnarray}
\label{eq:X_const}
X_n &=& \left(X_n^\c{L},X_n^\c{M} \right) = \left(U_n^\c{L},Z_n\ominus U_n^\c{M}\right) \in  \c{L}_n \times \c{M}_n^+ \\
Y_n &=& \left(U_n^\c{L},U_n^\c{M}\right) \in \c{L}_n \times \c{M}_n^+ 
\end{eqnarray}
where $\ominus$ represents the subtraction of modulo $M_n+1$.

Next, let us define the decoder $\psi_n^*$. Let $x_n = (x_n^\c{L},x_n^\c{M})\in \c{L}_n \times \c{M}_n^+$ and $y_n = (y_n^\c{L},y_n^\c{M})\in \c{L}_n \times \c{M}_n^+$ be the inputs to the decoder. Then, the decoder $\psi_n^*$ first checks whether $x_n^\c{L} = y_n^\c{L}$ holds or not. If $x_n^\c{L} \neq y_n^\c{L}$, the decoder judges that impersonation attack has occurred and outputs $\perp$. On the other hand, if $x_n^\c{L} = y_n^\c{L}$, the decoder computes $x_n^\c{M} \oplus y_n^\c{M}$, where $\oplus$ denotes the addition of modulo $M_n+1$. If $x_n^\c{M} \oplus y_n^\c{M}= M_n$, the decoder outputs $\perp$ since the decoding error occurs in such a case. Otherwise, the decoder outputs $\xi_n^{-1} (x_n^\c{M} \oplus y_n^\c{M})$ where $\xi^{-1}_n: \c{M}_n \rightarrow \c{T}_{\gamma_n}$ is the inverse map of $\xi_n$. Summarizing, the decoder $\psi_n^*$ is written as
\begin{eqnarray}
\psi_n^* ({x}_n,{y}_n)= 
\left\{
\begin{array}{cl@{}c@{}c@{}l@{}l}
\xi_n^{-1}(x_n^\c{M} \oplus y_n^\c{M}), &\mbox{if~} x_n^\c{L} = y_n^\c{L}&\mbox{~and~}& x_n^\c{M} \oplus y_n^\c{M} \neq M_n &\mbox{~are satisfied}\\
\bot, &\mbox{otherwise}&&&
\end{array}
\right.
\end{eqnarray}
and the acceptance region of $\psi_n^*$ is given by
\begin{eqnarray}
\c{A}_n = \{(x_n,y_n)\in{\cal X}_n \times {\cal Y}_n: x_n^\c{L} = y_n^\c{L}~\mbox{and}~x_n^\c{M} \oplus y_n^\c{M}\in \c{M}_n  \}.
\end{eqnarray}

Hereafter, we prove that the above sequence $\{(\varphi_n^*,\psi_n^*)\}_{n=1}^\infty$ realizes the optimal $(2,2)$--threshold scheme with correlation level $\ell$ that asymptotically attains all the bounds in (\ref{eq:direct1})--(\ref{eq:direct4}). It suffices to prove Claims \ref{claim:sizes}--\ref{claim:c_level} below. 

\begin{claim}\label{claim:sizes}
For an arbitrarily small $\gamma > 0$, the rates of $X_n$, $Y_n$ and $U_n$ cannot be  less than $H(S)+\ell-\gamma$ for all sufficiently large $n$, i.e., (\ref{eq:direct1})--(\ref{eq:direct3}) hold. 
\end{claim}
\bigskip
\begin{claim}\label{claim:success_p}
The limit inferior of the minimum exponent in the success probabilities of impersonation attacks is at least $\ell$, i.e., (\ref{eq:direct4}) holds. 
\end{claim}
\bigskip

\begin{claim}\label{claim:erorr}
The decoding error probability for the legitimate shares vanishes as $n$ goes to infinity, i.e., (\ref{eq:asym_err}) holds. 
\end{claim}
\bigskip

\begin{claim}\label{claim:security}
For all $n \ge 1$, the $n$ source outputs $S^n$ are secure against the leakage from one of $X_n$ and $Y_n$, i.e., (\ref{eq:direct5}) holds. 
\end{claim}
\bigskip

\begin{claim}\label{claim:c_level}
The correlation level between $X_n$ and $Y_n$ equals to $\ell$, i.e., (\ref{eq:corr-level}) holds. 
\end{claim}
\bigskip

{\em Proof of Claim \ref{claim:sizes}:}  In order to evaluate the share rates and the randomness given by (\ref{eq:direct1})--(\ref{eq:direct3}), observe that 
\begin{eqnarray}
\nonumber
\log|\c{X}_n| &=& \log|\c{Y}_n| = \log|\c{U}_n| \\
\nonumber
&=& \log \{L_n(M_n+1)\}\\
\nonumber
&=& \log \left\{\lfloor 2^{n\ell} \rfloor (|\c{T}_{\gamma_n}|+1)  \right\}\\
\label{eq:sizes}
&\le& n\{H(S) + \ell + \gamma_n \}+1
\end{eqnarray}
where the last inequality follows from (\ref{eq:type}). Hence, it holds that 
\begin{eqnarray}
\label{eq:rates}
\frac{1}{n}\log|\c{X}_n| = \frac{1}{n}\log|\c{Y}_n| = \frac{1}{n}\log|\c{U}_n| \le 
H(S) + \ell + \gamma_n + \frac{1}{n}.
\end{eqnarray}
Taking the limit superior of both sides in (\ref{eq:rates}), Claim \ref{claim:sizes} is established. \QED

\bigskip
{\em Proof of Claim \ref{claim:success_p}:}  We evaluate $P^X_n$ in the following way:
\begin{eqnarray}
\nonumber
 P_n^X &=& \max_{P_{\o{X}_n}}\pr \left\{(\o{X}_n,Y_n) \in \c{A}_n \right\}\\
\nonumber
&=& \max_{P_{\o{X}_n}}\pr \left\{\o{X}_n^\c{L} = Y_n^\c{L}~\mbox{and}~ \o{X}_n^\c{M} \oplus Y_n^\c{M} \in \c{T}_{\gamma_n}\right\}\\
\nonumber
&\le &\max_{P_{\o{X}_n}}\pr \left\{\o{X}_n^\c{L} = Y_n^\c{L}\right\}=\max_{P_{\o{X}_n^\c{L}}}\pr \left\{\o{X}_n^\c{L} = U_n^\c{L}\right\}\\
\nonumber
&\stackrel{\rm (a)}{=}&\max_{P_{\o{X}_n^\c{L}}}\sum_{ x_n^\c{L}  \in \c{L}_n}P_{\o{X}_n^\c{L}}(x_n^\c{L}) P_{U_n^\c{L}}(x_n^\c{L})\\
&\stackrel{\rm (b)}{=}&\frac{1}{L_n}\max_{P_{\o{X}_n^\c{L}}}\sum_{ x_n^\c{L}  \in \c{L}_n}P_{\o{X}_n^\c{L}}(x_n^\c{L})=\frac{1}{L_n}
\end{eqnarray}
where $\o{X}_n \DEF (\o{X}_n^\c{L},\o{X}_n^\c{M}) \in \c{L}_n \times \c{M}_n^+$ and 
$Y_n \DEF (Y_n^\c{L},Y_n^\c{M}) =(U_n^\c{L},U_n^\c{M}) \in \c{L}_n \times \c{M}_n^+$, and the marked equalities follow from the following reasons:
\begin{itemize}
\item[(a)]
$\o{X}_n^\c{L}$ and $U_n^\c{L}$ are independent.
\item[(b)]
$P_{U_n^\c{L}}(x^\c{L}_n)=1/L_n$ holds for all $x_n^\c{L}\in\c{L}_n$.
\end{itemize}
Similarly, noticing the fact that $X_n^\c{L}=U_n^\c{L}$, we also have $ P_n^Y \le 1/L_n$, and therefore, we conclude that 
\begin{eqnarray}
\liminf_{n\rightarrow \infty} \min\left\{-\frac{1}{n}\log  P_n^X,-\frac{1}{n}\log  P_n^Y\right\}\ge\liminf_{n\rightarrow \infty} \frac{1}{n}\log L_n = \ell.
\end{eqnarray}
\QED

{\em Proof of Claim \ref{claim:erorr}:} Since every legitimate pair $(x_n,y_n) \in \c{A}_n$ of shares is decoded by $\varphi_n^*$ without error, the decoding error happens only if the decoder $\psi_n^*$ outputs $\perp$ for a pair of  legitimate shares $(x_n,y_n)$. Hence, the decoding error probability $P_n^e$ can be written as 
\begin{eqnarray}
\nonumber
P_n^e&=& \pr\{\psi_n^*(X_n,Y_n) = \perp\} \\
\nonumber
&=& \pr \{\xi^+_n(S^n) = M_n\} \\
\nonumber
&=&\pr\{S^n \not\in \c{T}_{\gamma_n} \}.
\end{eqnarray}
Therefore, it follows from (\ref{eq:lln}) that
$\lim_{n \rightarrow \infty}P_n^e = 1 - \lim_{n \rightarrow \infty} \pr \{S^n \in \c{T}_{\gamma_n}\} =0$.  \QED

\bigskip
{\em Proof of Claim \ref{claim:security}:} First, we note that $Z_n$ and $X_n^\c{M}=Z_n \ominus U_n^\c{M}$ are independent because of  non-negativity of the mutual information and 
\begin{eqnarray}
\nonumber
I(Z_n;Z_n \ominus U_n^\c{M}) &=& H(Z_n) + H(Z_n \ominus U_n^\c{M}) - H(Z_n, Z_n \ominus U_n^\c{M}) \\
\nonumber
&=& H(Z_n) + H(Z_n \ominus U_n^\c{M}) - H(Z_n,U_n^\c{M}) \\
\nonumber
&=& H(Z_n \ominus U_n^\c{M}) - H(U_n^\c{M}) \\
&\le& 0
\end{eqnarray}
where the last inequality holds because $Z_n \ominus U_n^\c{M}\in \c{M}_n^+$ and $U_n^\c{M}$ is subject to the uniform distribution on $\c{M}_n^+$. Hence, $Z_n$ and $X_n=(X_n^\c{L},X_n^\c{M})$ are also independent because 
\begin{eqnarray}
\nonumber
I(Z_n;X_n) &=& I(Z_n;X_n^\c{L}X_n^\c{M}) \\
\nonumber
&=& I(Z_n;X_n^\c{M}) +I(Z_n;X_n^\c{L}|X_n^\c{M}) \\
\label{eq:security_verif}
&=&0
\end{eqnarray}
where the last equality follows since $I(Z_n;X_n^\c{M})=0$, and $Z_n$, $X_n^\c{M}$, and $X_n^\c{L}$ form a Markov chain in this order. 

In order to show (\ref{eq:direct5}), it is sufficient to prove that $ I(S^n;X_n) =0$ for all $n \ge 1$ because $I(S^n;Y_n) =0$ for any $n \ge 1$ trivially holds from the fact that $S^n$ and $Y_n=(U_n^\c{L},U_n^\c{M})$ are independent. In addition, $I(S_n;X_n)=0$ is established from $I(S^n;X_n) \le I(Z_n;X_n)=0$ which is obtained by the information processing inequality \cite[Theorem 2.8.1]{CT-06} for a Markov chain $S_n \rightarrow Z_n\rightarrow X_n$, and recalling (\ref{eq:security_verif}). \QED
\bigskip

{\em Proof of Claim \ref{claim:c_level}:} The correlation level can be evaluated as follows. Note that the mutual information of shares $X_n$ and $Y_n$ satisfies 
\begin{eqnarray}
\nonumber
I(X_n;Y_n) &=& I(X_n^\c{L}X_n^\c{M};Y_n^\c{L}Y_n^\c{M}) \\
\nonumber
& =& I(U_n^\c{L}X_n^\c{M};U_n^\c{L}U_n^\c{M}) \\
\nonumber
&=& H(U_n^\c{L}X_n^\c{M})  -H(X_n^\c{M}|U_n^\c{L}U_n^\c{M})-H(U_n^\c{L}|X_n^\c{M}U_n^\c{L}U_n^\c{M})\\
\nonumber
&\stackrel{\rm (c)}{=}& H(U_n^\c{L}) + H(X_n^\c{M}) - H(X_n^\c{M}|U_n^\c{M})\\
&\stackrel{\rm (d)}{=}& H(U_n^\c{L}) + H(X_n^\c{M}) - H(Z_n)
\label{eq:mutual_I}
\end{eqnarray}
where the marked equalities hold because of the following reasons:
\begin{itemize}
\item[(c)] $U_n^\c{L}$ and $ X_n^\c{M}$ are independent, and $U_n^\c{L}$, $U_n^\c{M}$, and $X_n^\c{M}$ form a Markov chain in this order. 
\item[(d)] It follows that 
$H(X_n^\c{M}|U_n^\c{M}) =  H\left(Z_n \ominus U_n^\c{M}|U_n^\c{M}\right) =  H\left(Z_n|U_n^\c{M}\right)= H\left(Z_n\right)$ 
due to the independence of $S^n$ and $U_n$.
\end{itemize}
Hereafter, we evaluate the terms on the right hand side of (\ref{eq:mutual_I}). It is easy to see that 
\begin{eqnarray}
\label{eq:eval_U1}
H(U_n^\c{L}) = \log L_n = \log \lfloor 2^{n\ell} \rfloor.
\end{eqnarray}
The second term in the right hand side of (\ref{eq:mutual_I}) can be evaluated as 
\begin{eqnarray}
\nonumber
H(X_n^\c{M}) &\le& \log (M_n+1) \\
\nonumber
&=& \log (|\c{T}_{\gamma_n}| +1) \\
\label{eq:tech}
&\le& n \{ H(S) + \gamma_n\}+1
\end{eqnarray}
where the last inequality follows from (\ref{eq:type}). In order to evaluate the last term on the right hand side of (\ref{eq:mutual_I}), we set $\delta_n = \pr \{\xi^+_n(S^n) = M_n\}= \pr\{S^n \not\in \c{T}_{\gamma_n}\}$. Clearly, $\lim_{n \rightarrow \infty} \delta_n = 0$ from (\ref{eq:lln}). Since the map $\xi_n:\c{T}_{\gamma_n} \rightarrow \c{M}_n$ is bijective, we have
\begin{eqnarray}
\nonumber
H(Z_n)&=&H(\xi^+_n(S^n))\\
\nonumber
&=& \sum_{s^n \in \c{T}_{\gamma_n}} P_{S^n}(s^n) \log \frac{1}{P_{S^n}(s^n)} + \delta_n \log \frac{1}{\delta_n}\\
\nonumber
&\ge&  \sum_{s^n \in \c{T}_{\gamma_n}} P_{S^n}(s^n) n\{H(S) -\gamma_n\} - \delta_n \log \delta_n\\
\label{eq:58}
&=& (1-\delta_n) n\{H(S) -\gamma_n\} - \delta_n \log \delta_n
\end{eqnarray}
where the inequality holds because of (\ref{eq:type}). Hence, we have from (\ref{eq:tech}) and (\ref{eq:58}) that 
\begin{eqnarray}
\label{eq:upper_X&Z}
H(X_n^\c{M}) - H(Z_n) \le 
n\delta_n H(S) + n(2-\delta_n) \gamma_n +\delta_n \log \delta_n+1.
\end{eqnarray}
On the other hand, it is easy to see with the same reason for the equality (d) in (\ref{eq:mutual_I}) that 
\begin{eqnarray}\label{eq:lower_X&Z}
H(X_n^\c{M}) - H(Z_n) \ge H(X_n^\c{M}|U_n^\c{M}) - H(Z_n) = 0. 
\end{eqnarray}
Summarizing, we have from (\ref{eq:mutual_I}), (\ref{eq:eval_U1}), (\ref{eq:upper_X&Z}), and (\ref{eq:lower_X&Z}) that 
\begin{eqnarray}\label{eq:eval_I}
\frac{1}{n} \log \lfloor 2^{n\ell} \rfloor \le \frac{1}{n}I(X_n;Y_n) \le \frac{1}{n} \log \lfloor 2^{n\ell} \rfloor + 
\delta_nH(S) + (2-\delta_n) \gamma_n + \frac{1}{n} \left(\delta_n \log \delta_n+1\right).
\end{eqnarray}
By taking the limit of both sides of (\ref{eq:eval_I}) and noticing that $\lim_{n \rightarrow \infty} \gamma_n = \lim_{n \rightarrow \infty} \delta_n =0$, we have 
\begin{eqnarray}
\label{eq:eval_U2}
\lim_{n \rightarrow \infty} \frac{1}{n} I(X_n;Y_n) = \ell. 
\end{eqnarray}
\QED

Since Claims \ref{claim:sizes}--\ref{claim:c_level} are verified, Theorem \ref{thm:direct} is proved. \QED 
\bigskip
\begin{rem}
The claim of Theorem~\ref{thm:direct} is valid for the class of stationary 
ergodic sources if the entropy $H(S)$ in Theorem \ref{thm:direct} is replaced 
with the entropy rate 
$H \DEF \lim_{n \to \infty} H(S^n)/n$. 
This fact is obtained by a slight modification 
of the proof of Theorem~\ref{thm:direct} followed by the diagonal line argument 
\cite[Theorem 1.8.2]{H-03}. First, by the asymptotic equipartition property \cite[Theorem 3.1.2]{CT-06}, 
we have 
\begin{eqnarray}
\lim_{n \to \infty} \Pr\{S^n \in \c{T}_{n,\gamma}\} = 1
\end{eqnarray}
for any constant $\gamma> 0$, where 
\begin{eqnarray}
\c{T}_{n,\gamma} \DEF \left\{
s^n \in \c{S}^n \,:\, \biggl|
\frac{1}{n} \log \frac{1}{P_{S^n}(s^n)} - H \biggr| \le \gamma
\right\}
\end{eqnarray}
and $H$ denotes the entropy rate of the source. 
We construct an encoder $\varphi_{n,\gamma}^\ast$ 
and a decoder $\psi_{n,\gamma}^\ast$ in the same way as 
in the proof of Theorem \ref{thm:direct}. 
It is easily checked that 
$\{(\varphi_{n,\gamma}^\ast,\psi_{n,\gamma}^\ast)\}_{n=1}^\infty$ 
asymptotically realizes the $(2,2)$--threshold scheme. In addition, by the same argument with (\ref{eq:rates}) and (\ref{eq:eval_I}), $\{(\varphi_{n,\gamma}^\ast,\psi_{n,\gamma}^\ast)\}_{n=1}^\infty$  satisfies 
\begin{equation}
\frac{1}{n} \log |\c{X}_n| 
= \frac{1}{n} \log |\c{Y}_n| 
= \frac{1}{n} \log |\c{U}_n| 
\le H + \ell + \gamma + \frac{1}{n}
         \label{eq:rate}
\end{equation}
and 
\begin{equation}
\frac{1}{n} \log \lfloor 2^{n\ell}\rfloor 
\le \frac{1}{n}I(X_n;Y_n) 
\le 
\frac{1}{n} \log \lfloor 2^{n\ell}\rfloor 
+ \delta_{n,\gamma} H + (2 - \delta_{n,\gamma}) \gamma 
+ \frac{1}{n}(\delta_{n,\gamma} \log \delta_{n,\gamma} + 1)
                 \label{eq:key}
\end{equation}
where $\delta_{n,\gamma} \DEF 
\Pr\{S^n \notin \c{T}_{n,\gamma}\} \to 0$ as $n \to \infty$. 
Note that (\ref{eq:key}) implies that 
\begin{equation}
\left|\,\frac{1}{n}I(X_n;Y_n) - \ell \,\right| \le 3 \gamma \quad 
\mbox{for all $n \ge N_0(\gamma)$}. 
         \label{eq:view}
\end{equation}

We now fix a sequence $\{\gamma_m\}_{m=1}^\infty$ satisfying 
$\gamma_0 > \gamma_1 > \cdots > \gamma_m > \cdots > 0$ arbitrarily 
and define $N_0 = 1$ and $N_m,\,m=1,2,\ldots,$ as the minimum integer 
$N$ satisfying 
$|\,I(X_n;Y_n)/n - \ell \,| \le 3 \gamma_m$ for all $n \ge N$. 
Obviously, $\{N_m\}_{m=1}^\infty$ is monotone nondecreasing. 
We define 
$(\varphi_{n}^\ast,\psi_{n}^\ast)$ as 
$(\varphi_{n,\gamma_0}^\ast,\psi_{n,\gamma_0}^\ast)$ 
for each $1 \le n < N_1$ and 
$(\varphi_{n,\gamma_m}^\ast,\psi_{n,\gamma_m}^\ast)$ 
for each $N_m \le n < N_{m+1},\,m=1,2,\ldots$. 
Then, in view of (\ref{eq:rate}), (\ref{eq:view}) and 
$\gamma_m \downarrow 0$ as $m \to \infty$, we can conclude that 
$\{(\varphi_n^\ast,\psi_n^\ast)\}_{n=1}^\infty$ satisfies 
\begin{eqnarray}
\limsup_{n \to \infty}\frac{1}{n}\log |\c{X}_n| 
= \limsup_{n \to \infty}\frac{1}{n}\log |\c{Y}_n| 
= \limsup_{n \to \infty}\frac{1}{n}\log |\c{U}_n| \le H + \ell
\end{eqnarray}
and 
\begin{eqnarray}
\lim_{n \to \infty} \frac{1}{n} I(X_n;Y_n) = \ell. 
\end{eqnarray}
\end{rem}
\bigskip

\section{Another Optimal Scheme Using Symbolwise Encoding}\label{sec:simpler}

In Section \ref{sec:main}, we have shown by using blockwise coding that the sequence $\{(\varphi_n^*,\psi_n^*)\}_{n=1}^\infty$ of an encoder and a decoder realizes the asymptotically optimal $(2,2)$--threshold scheme with correlation level $\ell$. In addition, $\{(\varphi_n^*,\psi_n^*)\}_{n=1}^\infty$ also attains the maximum exponent in the success probabilities of impersonation attack which is given by $\ell$. In this section, by using a symbolwise encoding, we give a simple construction of $\{(\varphi_n^*,\psi_n^*)\}_{n=1}^\infty$ that realizes the asymptotically optimal $(2,2)$--threshold scheme with correlation level $\ell$ and the exponent in the success probability of impersonation attacks equals to $\ell$. In this construction, we use a pair $(f,g)$ of an encoder $f$ and a decoder $g$ for a $(2,2)$--threshold scheme for a single source output $S$.  In addition,  a one-sided test is used to detect the impersonation attacks. 

Let $S,U,X$ and $Y$ be random variables of a secret, a random number, and two shares taking values  in finite sets $\c{S},\c{U},\c{X}$ and $\c{Y}$ respectively. For a non-negative number $\ell$, we first define a pair $(f,g)$ of an encoder $f$ and a decoder $g$ for a $(2,2)$--threshold scheme with correlation level $\ell$. That is, the encoder $f:\c{S} \times \c{U} \rightarrow \c{X} \times \c{Y}$ is defined to be a deterministic map satisfying 
\begin{eqnarray}
\label{eq:security3}
H(S|X) &=& H(S|Y) =H(S) \\
\label{eq:decoding}
H(S|XY) &=& 0
\end{eqnarray}
in addition to 
\begin{eqnarray}\label{eq:corr-level2}
I(X;Y) = \ell
\end{eqnarray}
where shares $X$ and $Y$ are determined by $(X,Y) = f(S,U)$. Note that (\ref{eq:security3}) and (\ref{eq:decoding}) are the ordinary requirements for $(2,2)$--threshold schemes, i.e., (\ref{eq:security3}) guarantees that any information of $S$ does not leak from either one of the shares, and (\ref{eq:decoding}) implies that the secret $S$ can be decoded from $X$ and $Y$ without error. Hence, let $g: \c{X} \times \c{Y} \rightarrow {\cal S} \cup \{\lambda \}$ be a decoder corresponding to $f$ and satisfying $g(x,y)=\lambda$ for every $(x,y)\in \c{X} \times \c{Y}$ that does not belong to the range of $f$. Furthermore, (\ref{eq:corr-level2}) means that the correlation level of $X$ and $Y$ generated by the encoder $f$ is equal to $\ell$. We say that a pair $(f,g)$ of an encoder $f$ and a decoder $g$ realizes a $(2,2)$--threshold scheme with correlation level $\ell$ in the non-asymptotic sense if $(f,g)$ satisfies (\ref{eq:security3})--(\ref{eq:corr-level2}). In addition, it is shown in \cite{BSV-ipl98} that 
\begin{eqnarray}\label{eq:require}
\min\left\{\,|\c{X}|,|\c{Y}|,|\c{U}|\,\right\} \ge |\c{S}|
\end{eqnarray}
must be satisfied for any encoder of $(2,2)$--threshold schemes satisfying (\ref{eq:security3}) and (\ref{eq:decoding}). Hence, we also impose (\ref{eq:require}) on $f$ in addition to (\ref{eq:security3})--(\ref{eq:corr-level2}). 

In this setting, we define an encoder $\varphi_n^*: \c{S}^n \times \c{U}^n  \rightarrow \c{X}^n \times \c{Y}^n$ as the repeated application of $f: \c{S}\times \c{U}\rightarrow \c{X} \times \c{Y}$ to $(S_i,U_i)$, $i=1,2,\ldots,n$, which can be written as
\begin{eqnarray}\label{eq:const_phi}
\varphi_n^*(s^n,u^n) \DEF  f(s_1,u_1)f(s_2,u_2)\cdots f(s_n,u_n)
\end{eqnarray}
where $s^n \DEF  s_1s_2\cdots s_n\in\c{S}^n$ and $u^n \DEF  u_1u_2 \cdots u_n\in\c{U}^n$ are $n$ secrets and $n$ random numbers, respectively. Hence, the two shares $X^n=X_1 X_2 \cdots X_n \in \c{X}^n$ and $Y^n=Y_1Y_2 \cdots Y_n \in  \c{Y}^n$ are i.i.d.\ copies of $X$ and $Y$, respectively, where $(X_i,Y_i)=f(S_i,U_i)$. 

Furthermore, we define 
\begin{eqnarray}
\label{eq:auth-region}
\c{A}_n^*&=&\biggl\{
(x^n,y^n) \in \c{X}^n \times \c{Y}^n :\left.
\frac{1}{n} \log \frac{P_{XY}(x^n,y^n)}{P_{X^n}(x^n)P_{Y^n}(y^n)} > I(X;Y) - \gamma_n
\right\}
\end{eqnarray}
where $\gamma_n$ is an arbitrary sequence of positive integers $\{\gamma_n\}_{n=1}^\infty$ satisfying $\lim_{n \rightarrow \infty}\gamma_n=0$ and $\lim_{n \rightarrow \infty}\sqrt{n} \gamma_n=\infty$. Then, legitimate shares belong to $\c{A}_n^*$ with high probability if $n$ is sufficiently large since 
\begin{eqnarray}\label{eq:acceptance_prob}
\lim_{n\rightarrow\infty} \pr\{(X^n,Y^n) \in \c{A}_n^*\}=1
\end{eqnarray}
holds from the law of large numbers. Hence, we regard the received shares as legitimate if they belong to $\c{A}_n^*$, and decode them by the decoder $g_n$ corresponding to the encoder $\varphi_n^*$ in (\ref{eq:const_phi}), where $g_n$ can be written as 
\begin{eqnarray}
\label{eq:g_n}
g_n(x^n,y^n)\DEF g(x_1,y_1)g(x_2,y_2)\cdots g(x_n,y_n).
\end{eqnarray}
In addition, the decoder $\psi_n^*:\c{X}^n \times \c{Y}^n \rightarrow \c{S}^n \cup \{\perp\}$ is defined by 
\begin{eqnarray}\label{eq:decoder2}
\psi_n^* ({x}^n,{y}^n) =\left\{
\begin{array}{c@{}l}
g_n ({x}^n,{y}^n), &~\mbox{if}~({x}^n,{y}^n) \in \c{A}_n^*\\  
\perp, &   ~\mbox{otherwise}
\end{array} 
\right.
\end{eqnarray}
where $\perp$ means that the impersonation attack, i.e., {\sf (a1)} or {\sf (a2)} in Section \ref{sect:PS}, is detected.

According to (\ref{eq:acceptance_prob}), every $(x^n,y^n) \in \c{A}_n^*$ satisfies $P_{X^nY^n}(x^n,y^n) > 0$, which is equivalent to $P_{XY}(x_i,y_i)>0$, i.e., $g(x_i,y_i) \neq \lambda$, for all $i=1,2,\ldots,n$. Hence, for every $(x^n,y^n) \in \c{A}_n^*$, there uniquely exists $s^n \in \c{S}^n$ that satisfies $g_n(x^n,y^n)=s^n$ whether the received shares $x^n$ and $y^n$ are legitimate or not. Furthermore, if the pair of shares $(x^n,y^n)\in \c{A}^*_n$ is legitimate, the secret is reproduced without error due to the definitions of $f$ and $\varphi_n^*$. More precisely, $\psi^*_n(x^n,y^n)=g_n(x^n,y^n)=s^n$ holds for every $u^n \in \c{U}^n$, $s^n \in \c{S}^n$, $x^n \in \c{X}^n$ and $y^n \in \c{Y}^n$  satisfying $\varphi^*_n(s^n,u^n) = (x^n,y^n)\in \c{A}_n^*$.

The above sequence $\{(\varphi_n^*, \psi_n^*)\}_{n=1}^\infty$ defined by (\ref{eq:const_phi})  and (\ref{eq:decoder2}) realizes an asymptotic $(2,2)$--threshold scheme with correlation level $\ell$. 

\bigskip
\begin{thm}\label{thm:simple_direct}
Let $(f,g)$ be any pair of an encoder and a decoder that realizes a $(2,2)$--threshold scheme with correlation level $\ell$  in the non-asymptotic sense. Then, the sequence $\{(\varphi_n^*, \psi_n^*)\}_{n=1}^\infty$ defined by (\ref{eq:const_phi})  and (\ref{eq:decoder2}) satisfies for all $n \ge 1$ that 
\begin{eqnarray}
\label{eq:th3-1}
P_n^e & = & \pr \{(X^n,Y^n) \not\in \c{A}_n^*\} \\
\label{eq:th3-2}
H(S^n|X^n) &=& H(S^n|Y^n) = H(S^n) \\
\label{eq:th3-3}
I(X^n;Y^n) &=& n\ell
\end{eqnarray}
which obviously realizes an asymptotic $(2,2)$--threshold scheme with correlation level $\ell$.  In addition, this $\{(\varphi_n^*, \psi_n^*)\}_{n=1}^\infty$ satisfies (\ref{eq:direct4}). 
\end{thm}
\bigskip

{\em Proof of Theorem \ref{thm:simple_direct}:} First, we prove (\ref{eq:th3-1}). 
If there is a one-to-one correspondence between $(s^n,u^n)$ and $(x^n,y^n)$, (\ref{eq:th3-1}) is obvious. We show that (\ref{eq:th3-1}) holds for any pair of $f$ and $g$ satisfying (\ref{eq:security3}) and (\ref{eq:decoding}) which does not guarantee the existence of such a one-to-one correspondence.

Define
\begin{eqnarray}
\c{D}_n^*(x^n,y^n) = \{(s^n,u^n): \varphi^*_n (s^n,u^n) = (x^n,y^n)~\mbox{and}~\psi^*_n(x^n,y^n)=s^n\}.
\end{eqnarray}
Recalling that $\psi^*_n(x^n,y^n)=g_n(x^n,y^n)=s^n$ for every $u^n \in \c{U}^n$, $s^n \in \c{S}^n$, $x^n \in \c{X}^n$ and $y^n \in \c{Y}^n$  satisfying $\varphi^*_n(s^n,u^n) = (x^n,y^n)\in \c{A}_n^*$, it holds for all $(x^n,y^n) \in \c{A}^*_n$ that 
\begin{eqnarray}
\nonumber
\c{D}_n(x^n,y^n) &=& \{(s^n,u^n): \varphi^*_n (s^n,u^n) = (x^n,y^n)~\mbox{and}~g_n(x^n,y^n)=s^n\}\\
\nonumber
 &=& \{(s^n,u^n): \varphi^*_n (s^n,u^n) = (x^n,y^n)\}\\
\label{eq:key_rv}
&\DEF& {\varphi^*_n}^{-1}(x^n,y^n)
\end{eqnarray}
where ${\varphi^*_n}^{-1}(x^n,y^n)$ means the inverse image of $(x^n,y^n)$. 

Next, we define 
\begin{eqnarray}
\c{D}_n = \{(s^n,u^n): \psi^*_n(\varphi_n^*(s^n,u^n) )=s^n\}. 
\end{eqnarray}
Then, since $\psi^*_n(x^n,y^n) = \perp$ for all $(x^n,y^n) \not\in \c{A}_n^*$ and $s^n$ is reproduced without error from every $(x^n,y^n)\in \c{A}^*_n$, we have 
\begin{eqnarray}
\nonumber
\c{D}_n &=& \bigcup_{(x^n,y^n) \in \c{A}_n^*} \c{D}_n (x^n,y^n)\\
\label{eq:join_D}
&=& \bigcup_{(x^n,y^n) \in \c{A}_n^*} {\varphi^*_n}^{-1} (x^n,y^n)
\end{eqnarray}
where the second equality follows from (\ref{eq:key_rv}). Furthermore, since $\varphi^*_n$ is deterministic, it is easy to see that 
\begin{eqnarray}
\label{eq:disjoint}
{\varphi^*_n}^{-1}(x^n,y^n)  \cap {\varphi^*_n}^{-1}(\t{x}^n,\t{y}^n) = \emptyset ~~
\mbox{for all}~~(x^n,y^n) \neq (\t{x}^n,\t{y}^n). 
\end{eqnarray}
From (\ref{eq:join_D}) and (\ref{eq:disjoint}), it is shown that $\{{\varphi^*_n}^{-1}(x^n,y^n)\}_{(x^n,y^n) \in \c{A}^*_n}$ is a partition of $\c{D}_n$. Therefore,  we have
\begin{eqnarray}
\nonumber
1-P^e_n &=& \sum_{(s^n,u^n)\in\c{D}_n} P_{S^nU^n}(s^n,u^n).\\
\nonumber
&=& \sum_{(x^n,y^n) \in \c{A}^*_n}\sum_{(s^n,u^n)\in {\varphi^*_n}^{-1} (x^n,y^n)} P_{S^nU^n}(s^n,u^n)\\
\nonumber
&=&  \sum_{(x^n,y^n) \in \c{A}_n^*}P_{X^nY^n}(x^n,y^n)\\
&=&  \pr\{(X^n,Y^n)  \in \c{A}^*_n \}
\end{eqnarray}
where the first equality comes from the definition of the decoding error probability and the third equality is due to the definition of $P_{X^nY^n}(\cdot,\cdot)$, i.e., $P_{X^nY^n}(x^n,y^n)=\sum_{(s^n,u^n)\in\c{S}^n \times\c{U}^n: \varphi^*_n (s^n,u^n) = (x^n,y^n)}P_{S^nE^n}(s^n,u^n)$. 
Hence, we obtain (\ref{eq:th3-1}). It is easy to see  from (\ref{eq:acceptance_prob}) that $P_n^e$ satisfies (\ref{eq:asym_err}) i.e., the decoding error probability of $\psi_n^*$ in (\ref{eq:decoder2}) vanishes as $n$ goes to infinity.

In order to establish Theorem \ref{thm:simple_direct}, it remains to show that $\{(\varphi_n^*, \psi_n^*)\}_{n=1}^\infty$ satisfies (\ref{eq:direct4}). Note that (\ref{eq:th3-2}) and (\ref{eq:th3-3}) clearly hold from (\ref{eq:security3}) and (\ref{eq:corr-level2}), respectively. To this end, we evaluate the success probability of the impersonation attack as follows:
\begin{eqnarray}
\nonumber
 P_n^X
\nonumber
&=&\max_{P_{\o{X}^n}}\pr \left\{(\o{X}^n,Y^n)\in\c{A}_n^*\right\}\\
\nonumber
&=&\max_{P_{\o{X}^n}}\sum_{(x^n,y^n)  \in\c{A}_n^* }P_{\o{X}^n}(x^n)P_{Y^n}(y^n)\\
\nonumber
&{\le}&\max_{P_{\o{X}^n}}\sum_{(x^n,y^n)  \in\c{A}_n^* }P_{\o{X}^n}(x^n)\frac{P_{X^nY^n}(x^n,y^n)}{P_{X^n}(x^n)} 2^{-n(\ell - \gamma_n)}\\
\nonumber
&\le&\max_{P_{\o{X}^n}}\sum_{(x^n,y^n) \atop \in\c{X}^n \times \c{Y}^n }P_{\o{X}^n}(x^n)\frac{P_{X^nY^n}(x^n,y^n)}{P_{X^n}(x^n)} 2^{-n(\ell - \gamma_n)}\\
\label{eq:auth}
&=& 2^{-n(\ell - \gamma_n)}
\end{eqnarray}
where the first inequality follows from (\ref{eq:auth-region}) which implies  
\begin{eqnarray}
\nonumber
P_{Y^n}({y}^n) 
&<& \frac{P_{X^nY^n}(x^n,y^n)}{P_{X^n}(x^n)} 2^{-n\left\{I(X;Y)- \gamma_n\right\}}\\
&=& \frac{P_{X^nY^n}(x^n,y^n)}{P_{X^n}(x^n)} 2^{-n(\ell - \gamma_n)} 
\end{eqnarray}
for any $(x^n,y^n) \in \c{A}_n^*$. Similarly, we have $ P_n^Y \le  2^{-n(\ell - \gamma_n)}$. Hence, we obtain (\ref{eq:direct4}) since $\lim_{n\rightarrow \infty}\gamma_n = 0$. \QED

Since Theorem \ref{thm:simple_direct} has been proved, we are now interested in a relation between the share rates and the correlation level attained by a pair $(f,g)$ of an encoder $f$ and a decoder $g$, which is given by the following claim:
\bigskip
 
\begin{claim}\label{claim:converse-rep}
Let $M$ and $M_S$ be arbitrary positive integers satisfying $M \ge M_S$. Then, there exists a pair ($f^*,g^*)$ of an encoder $f^*$ and a decoder $g^*$ for a $(2,2)$--threshold scheme with correlation level $\ell = \log M - H(S)$ in the non-asymptotic sense satisfying $|\c{X}| = |\c{Y}| = |\c{U}| =  M$ and $|\c{S}| = M_S$.  
\end{claim}
\bigskip

\begin{rem}
According to Claim \ref{claim:converse-rep}, the rates of shares and randomness are $\log |\c{X}| = \log |\c{Y}| = \log |\c{U}| = \log M = H(S) - \ell$, which coincides with the lower bounds of the rates given by (\ref{eq:direct1})--(\ref{eq:direct3}). Hence, the sequence $\{(\varphi_n^*,\psi^*_n)\}_{n=1}^\infty$ defined by (\ref{eq:const_phi}) and (\ref{eq:decoder2}) also achieves all the bounds in Theorem \ref{thm:direct}. Observe that the sequence of encoders $\{\varphi_n^*\}_{n=1}^\infty$ in this section is simpler than the the sequence of encoders presented in Theorems 2. For instance, $S^n$ cannot be encoded symbolwisely by the sequence of encoders in the proof of Theorem \ref{thm:direct} since the correlation of two shares is generated by the random variable $U_n^\c{L}$ in both shares contained in common. On the other hand, symbolwise encoding is possible by the sequence of encoders in this section since $X_i$ and $Y_i$ are correlated due to $f$ for every $i=1,2,\ldots, n$. Furthermore, such symbolwise encoding also enables us that $I(S_i;X_i) = I(S_i;Y_i) =0$ for every $i =1,2,\ldots,n$, which is stronger than the security condition given by (\ref{eq:direct5}) in Theorem \ref{thm:direct}.

However, we note that $M$, $M_S$ and the correlation level $\ell$ cannot be set arbitrarily in Claim \ref{claim:converse-rep} although they can be taken arbitrarily in Theorem \ref{thm:direct}, which is compensation for the simplicity.
\end{rem}
\bigskip

\begin{rem}\label{rem:ideality} 
In the {\sssc} in a non-asymptotic setup (e.g., \cite{MS-cacm81,KGH-it83,TW-jc88,OKS-siam06,CSV-ecrypt93,KOO-crypto95}), it is shown that any ideal secret sharing scheme cannot detect any forgery of shares with probability $1$. Furthermore, as is shown in \cite{BSV-ipl98}, we note that the ideal secret sharing schemes can be realized if and only if $|\c{X}|=|\c{Y}|=|\c{S}|$ and $S$ is uniformly distributed.

Similarly, in the asymptotic setup discussed in this section, it is impossible for any $(f,g)$ of an ideal $(2,2)$--threshold scheme to achieve $P^X_n$ and $P^Y_n$ with exponential order of $n$  because the correlation level  $\log M - H(S) =0$ is satisfied if and only if $M=|\c{S}|$ and $S$ is uniformly distributed. On the other hand, we note that $\ell$ is positive for arbitrary distribution of $S$ if $\min\left\{\,|\c{X}|,|\c{Y}|,|\c{U}|\,\right\} > |\c{S}|$. \QED
\end{rem}
\bigskip

{\em Proof of Claim \ref{claim:converse-rep}:} From (\ref{eq:require}), let us define
\begin{eqnarray}
\c{X}&=&\c{Y}=\c{U}=\{0,1,\ldots,M-1\} \\
\c{S}&=& \{0,1,\ldots, M_S -1\}
\end{eqnarray}
where $M \ge M_S$. Define the encoder $f^*:\c{S} \times \c{U} \rightarrow \c{X} \times \c{Y}$ for a secret $s\in\c{S}$ and a random number $u\in\c{U}$ as 
\begin{eqnarray}\label{eq:enc}
f^*(s,u) = (s \ominus u ,u)
\end{eqnarray}
where $\ominus$ denotes the subtraction of modulo $M$. Then, the corresponding decoder $g^*:
\c{X}\times \c{Y} \rightarrow \c{S} \cup \{\lambda\}$ can be written as 
\begin{eqnarray}
\label{eq:decode}
g^*(x,y)=\left\{
\begin{array}{clc}
x \oplus y, & \mbox{if~}  x \oplus y \in \c{S}\\
\lambda, & \mbox{otherwise}
\end{array}
\right.
\end{eqnarray}
where $\oplus$ represents the addition of modulo $M$.  Note that the secret $s$ can be decoded by $g^*$ without error, and hence, (\ref{eq:decoding}) is satisfied. Furthermore, we can check that a pair of the shares $(X,Y)$ is generated according to the conditional probability distribution 
\begin{eqnarray}
\label{eq:F}
P_{XY|S}(x,y|s)=
\left\{
\begin{array}{clc}
1/M, & \mbox{if~~} s= x \oplus y \in \c{S} \\
0, & \mbox{otherwise}
\end{array}
\right.
\end{eqnarray}
if we apply the encoder $f^*$ defined in (\ref{eq:enc}) to the secret $S$ with an arbitrary probability distribution $P_S(\cdot)$. Hence, the following discussion holds for an arbitrary distribution on $S$. This idea is based on the secret sharing scheme for non-uniform secret distribution studied in \cite{BSV-ipl98}. 

We show that (\ref{eq:security3}) is satisfied by $X$ and $Y$ generated by $f^*$. For every fixed $x\in\c{X}$ and $s\in\c{S}$, we can check that there exists a unique $y \in \c{Y}$,  satisfying $s=g^*(x,y)$. Hence, it holds from (\ref{eq:F}) that 
\begin{eqnarray}
\nonumber
P_{X|S}(x|s)&=&\sum_{y \in \c{Y}}P_{XY|S}(x,y|s)\\
\label{eq:g-property1}
&=&\frac{1}{M}
\end{eqnarray}
for every $(x,s) \in\c{X} \times \c{S}$. Then, we have 
\begin{eqnarray}
\nonumber
P_{X}(x) &=& \sum_{s \in \c{S}} P_{X|S}(x|s) P_S(s)\\
\nonumber
&=& \sum_{s \in \c{S}} \frac{1}{M}\cdot P_S(s) \\
\label{eq:g-property2}
&=& \frac{1}{M}.
\end{eqnarray}
From (\ref{eq:g-property1}) and (\ref{eq:g-property2}), it is shown that $S$ and $X$ are statistically independent. Similarly, it can be shown that $S$ and $Y$ are statistically independent, and hence, (\ref{eq:security3}) is proved.

The correlation level of $X$ and $Y$ generated by $f^*$ can be calculated as follows. We note that 
\begin{eqnarray}
\nonumber
H(XY) &\stackrel{\rm (e)}{=}& H(US)\\
\nonumber
& \stackrel{\rm (f)}{=}& H(U) +  H(S)\\
\label{eq:g-property3}
&=& \log M + H(S)
\end{eqnarray}
where the marked equalities (e) and (f) hold since 
\begin{itemize}
\item[(e)] there exists a bijection between $\c{U} \times \c{S}$ and $\c{X} \times \c{Y}$. 
\item[(f)] $U$ and $S$ are statistically independent. 
\end{itemize}
Therefore, we obtain from (\ref{eq:g-property2}) and (\ref{eq:g-property3}) that 
\begin{eqnarray}
\nonumber
I(X;Y) &=& H(X)+H(Y)-H(XY) \\
\nonumber
&=& 2\log M - \left\{\log M + H(S) \right\} \\
\label{eq:key-relation}
&=& \log M -  H(S).
\end{eqnarray}
Hence, it is shown that the pair $(f^*,g^*)$ of the encoder and the decoder actually realizes a $(2,2)$--threshold scheme with correlation level $\log M -H(S)$. 
\QED

\section{Conclusion}
This paper is concerned with coding theorems 
for a $(2,2)$--threshold scheme in the presence of 
an opponent who impersonates one of the participants. 
We have considered an asymptotic setup of the $(2,2)$--threshold scheme 
in which $n$ secrets from a memoryless source are encoded to two shares 
by using a uniform random number,  
and the two shares are decoded to the $n$ secrets 
with permitting negligible decoding error probability. We have investigated 
the minimum attainable rates of the two shares and 
the uniform random number, and the maximum exponents of
the probabilities of the successful impersonation 
from a Shannon-theoretic viewpoint.  
We have presented coding theorems for two cases of encoding, i.e., blockwise and symbolwise encoding. 

In the first case, 
we have considered the situation where the $n$ secrets 
are encoded blockwisely to two shares. 
We have defined the correlation level 
$\ell \ge 0$ of the shares as the limit of 
the normalized mutual information between the two shares. 
In the converse part it is shown that for any sequence 
$\{(\varphi_n,\psi_n)\}_{n=1}^\infty$ of pairs of an encoder 
$\varphi_n$ and a decoder $\psi_n$ that asymptotically realizes 
a $(2,2)$--threshold scheme with the correlation level $\ell$, 
none of the rates can be less than $H(S)+\ell$, 
where $H(S)$ denotes the entropy of the source, and 
the exponent of the probability of the successful impersonation 
cannot be less than $\ell$. In addition, we have shown 
the existence of a sequence $\{(\varphi_n^*,\psi_n^*)\}_{n=1}^\infty$ 
of pairs of an encoder $\varphi_n^*$ and a decoder $\psi_n^*$ 
that attains all the bounds given in the converse part. 
The obtained results can be easily extended to the case 
where the $n$ secrets are generated from a stationary ergodic source. 

In the second case, we have considered the situation 
where the $n$ secrets are encoded symbolwisely to two shares of 
length $n$ by repeatedly applying the encoder of 
an ordinary $(2,2)$--threshold scheme to the $n$ secrets. 
While the above converse part is valid in this setup, 
we can give another interesting decoder in the direct part. 
That is, we have shown that 
the impersonation by an opponent can be verified with probability 
close to one by verifying the joint typicality of the two shares. 
It turns out that these encoder and decoder also attain 
all the bounds in the converse part. 

\section*{Acknowledgment}
The authors would like to thank Prof.\ Hiroshi Nagaoka 
in the University of Electro-Communications, for his helpful comments. 
The work of M.\ Iwamoto is partially supported
by the MEXT Grant-in-Aid for Young Scientists (B) No.\ 20760236. 
The work of H.\ Koga is supported in part  
by Grant-in-Aid from the Telecommunications Advancement Foundation. 


\end{document}